\documentclass{appolb}
\usepackage{epsfig}

\begin{document}
\pagestyle{plain} \eqsec
\newcount\eLiNe\eLiNe=\inputlineno\advance\eLiNe by -1
\title{Nonmonotonic Behavior of Spatiotemporal Pattern Formation in
a Noisy Lotka-Volterra System \thanks{Presented at the $16^{th}$
Marian Smoluchowski Symposium on Statistical Physics: Fundamentals
and Applications, Zakopane, Poland, September 6-11, 2003}}%

\author{A.Fiasconaro\footnote{E-mail adress: afiasconaro@gip.dft.unipa.it},
D.Valenti, B.Spagnolo \address{INFM, Group of Interdisciplinary Physics and\\
Dipartimento di Fisica e Tecnologie Relative, Universit\`a di
Palermo \\ Viale delle Scienze - 90128 Palermo, Italy}} \maketitle

\begin{abstract}
The noise-induced pattern formation in a population dynamical
model of three interacting species in the coexistence regime is
investigated. A coupled map lattice of Lotka-Volterra equations in
the presence of multiplicative noise is used to analyze the
spatiotemporal evolution. The spatial correlation of the species
concentration as a function of time and of the noise intensity is
investigated. A nonmonotonic behavior of the area of the patterns
as a function of both noise intensity and evolution time is found.
\end{abstract}

\PACS{05.40.-a, 87.23.Cc, 89.75.Kd, 87.23.-n}

\section{Introduction}
The study of models of population dynamics is one of the topics
that in the last few years has revealed large interest ranging
from chaos to spatial organization, covering several branches of
physics and different disciplines such as biology, theoretical
ecology, oceanography and medicine. In particular the addition of
noise in such mathematical models can be useful to describe the
phenomenology in a realistic and relatively simple form. The
introduction of noise in real systems gives rise to non trivial
effects, modifying sometimes in an unexpected way the
deterministic dynamics. Examples of noise induced phenomena are
stochastic resonance \cite{benzi,valenti}, noise-enhanced
stability \cite{Agu}, temporal oscillations and noise-induced
pattern formation \cite{Gar,Sancho,Katja}. Biological complex
systems can be modelled as open systems in which interactions
between the components are nonlinear and a noisy interaction with
the environment is present \cite{Ciuchi}. The dynamics of
interacting biological species can be successfully described by
means of a set of Lotka-Volterra equations with the addition of a
multiplicative noise and a diffusive term to take into account
spatial extension of the ecosystem. Recently it has been found
that nonlinear interaction and the presence of multiplicative
noise can give rise to pattern formation in population dynamics of
spatially extended systems \cite{Spa1,Spa2}. The model studied in
this work concerns the ecosystem composed of three interacting
species: two competing preys and one predator. The area of the
patterns of maximum density has been quantitatively evaluated,
finding a nonmonotonic behavior as a function of time and as a
function of the noise intensity. The site correlation between the
species concentration as a function of time and of the noise
intensity is also investigated.
\section{The model}
The starting point of our study is the simple discrete set of
generalized Lotka-Volterra equations describing a population of
two preys and one predator. The dynamics of our spatially
distributed system is therefore described by the following model
of coupled map lattice (CML) \cite{cml}
\begin{eqnarray}
x_{i,j}^{n+1} & = & \mu x_{i,j}^n (1 - \nu x_{i,j}^n-\beta^n
y_{i,j}^n-\alpha z_{i,j}^n)+\sqrt{q} x_{i,j}^n X_{i,j}^n + D\sum_p
(x_{p}^n-x_{i,j}^n),
\nonumber \\
y_{i,j}^{n+1} & = & \mu y_{i,j}^n (1 - \nu y_{i,j}^n-\beta^n
x_{i,j}^n-\alpha z_{i,j}^n)+\sqrt{q} y_{i,j}^n Y_{i,j}^n +D\sum_p
(y_{p}^n-y_{i,j}^n),
\nonumber \\
z_{i,j}^{n+1} & = & \mu_z z_{i,j}^n
[-1+\gamma(x_{i,j}^n+y_{i,j}^n)] + \sqrt{q}z_{i,j}^n Z_{i,j}^n +
D\sum_p (z_{p}^n-z_{i,j}^n),
 \label{eqset}
\end{eqnarray}
where $x_{i,j}^n$, $y_{i,j}^n$ and $z_{i,j}^n$ are respectively
the densities of preys $x$, $y$ and the predator $z$ in the site
$(i,j)$ at the time step $n$. Here $\alpha$ and $\gamma$ are the
interaction parameters between preys and predator, $X$, $Y$ and
$Z$ are the white Gaussian noise variables with
 \begin{equation}
   \langle X(t)\rangle =  \langle Y(t)\rangle =  \langle Z(t)\rangle = 0,
 \end{equation}
 \begin{equation}
   \langle X(t) X(t + \tau)\rangle  = \langle Y(t) Y(t + \tau)\rangle  =
   \langle Z(t) Z(t + \tau)\rangle = \delta(\tau),
 \end{equation}
 \begin{equation}
   \langle X(t) Y(t')\rangle  = \langle X(t) Z(t')\rangle  =
   \langle Y(t) Z(t')\rangle = 0 \;\;\; \forall \; t,t'
 \end{equation}
$q$ is the noise intensity, $D$ is the diffusion coefficient,
$\mu$ and $\mu_z$ are scale factors. $\sum_{p}$ indicates the sum
over the four nearest neighbors in the map lattice. The
multiplicative noise in the above equations models the interaction
between the species and the environment. The boundary conditions
have been established in such a way that no interaction is present
out of lattice. This means that for the four corner sites we have
only two interactions and for the other 4x98 line-confined sites
the number of interactions is three.

We analyze the transient dynamics of the system with a time
varying interaction parameter $\beta(t)$ between the two preys
\begin{equation}
 \beta(t)=1 + \epsilon + \eta cos(\omega t),
 \label{betat}
\end{equation}
due to the environment temperature. Here $\eta = 0.2$, $\omega =
\pi 10^{-3}$ and $\epsilon=-0.1$. The interaction parameter
$\beta(t)$ oscillates around the critical value $\beta_c=1$ in
such a way that the dynamical regime of Lotka-Volterra model for
two competing species changes from coexistence of the two preys
($\beta<1$) to exclusion of one of them ($\beta>1$) \cite{Baz}.
The parameters used in our simulations are $\mu = 2$; $\nu = 1$;
$\alpha = 0.03$; $\mu_z = 0.02$; $\gamma = 205$ and $D = 0.1$. The
noise intensity $q$ varies between $10^{-12}$ and $10^{-2}$. With
this choice of parameters the interspecies competition among the
two prey populations is stronger compared to the intraspecies
competition (preys-predator), and, therefore, both prey
populations can stably coexist in the presence of the predator
\cite{Baz}.

 By considering the influence of noise and spatial diffusion, we
analyze the noise-induced pattern formation. Specifically the time
evolution of the area of the patterns and the correlation $r$ over
the grid between two species, as a function of the noise
intensity, are analyzed. The quantitative calculations of the site
correlation between a couple of species in the lattice have been
done using the following formula
\begin{equation}
r^n = \frac{\sum_{i,j}^N (w_{i,j}^n - \bar{w}^n) (k_{i,j}^n -
\bar{k}^n)}{\left[\sum_{i,j}^N (w_{i,j}^n - \bar{w}^n)^2
\sum_{i,j}^N (k_{i,j}^n - \bar{k}^n)^2 \right]^{1/2}} \label{r}
\end{equation}
where $N$ is the number of sites in the grid, the symbols $w^n,
k^n$ represent one of the three species $x, y, z$, and
$\bar{w}^n,\bar{k}^n$ represent the mean values of the
concentration of the species in all the lattice at the step $n$.
From the definition (\ref{r}) it follows
\begin{equation}
 -1 \leq r^n \leq 1.
\end{equation}
We calculate the correlation between the two preys and between
preys and predator.

\section{Results and Comments}
The bidimensional spatial grid considered is composed by 100x100
sites in $(x,y)$ plane. The calculations have been done for
various noise intensities and at different steps of the iteration
process. To quantify our analysis we consider only the maximum
patterns, defined as the ensemble of adjoining sites in the
lattice for which the density of the species belongs to the
interval $[3/4 \; max, max]$, where $max$ is the absolute maximum
of density in the specific grid.
\begin{figure}[htbp]
 \begin{center}
  \includegraphics[height=4.6cm]{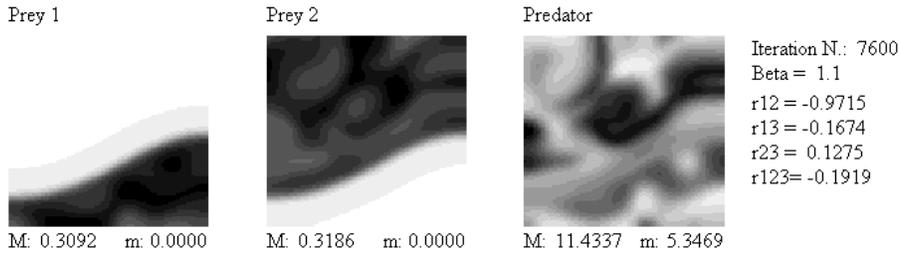}
  \vskip -0.4cm
\caption{Spatial configuration of the species obtained after long
time evolution in noiseless dynamics and in the exclusion regime
between preys. The parameter set is: $\epsilon = 0.1, \beta = 1.1,
\eta = 0, q = 0, D = 0.1, \mu = 2, \nu = 1, \alpha = 0.03, \mu_z =
0.02, \gamma = 205.$ The initial conditions are random with a
Gaussian distribution, with mean values
$\bar{x}(0)=\bar{y}(0)=\bar{z}(0)= 0.25$ and variance $\sigma_o =
0.1$. Here $r_{12}, r_{13}, r_{23}$ and $r_{123}$ are respectively
the site correlations between: (i) preys, (ii) prey 1 and
predator, (iii) prey 2 and predator, and (iv) predator and both
preys.}
  \label{pat1}
 \end{center}%
\end{figure}
For each spatial distribution, in a temporal step and for a given
noise intensity value, the following quantities have been
evaluated referring to the maximum pattern (MP): mean area of the
various MPs found in the lattice and spatial correlation $r$
between two preys, and between preys and predator.
\subsection{Deterministic Analysis}
\begin{figure}[htbp]
 \begin{center}
  \includegraphics[height=8cm]{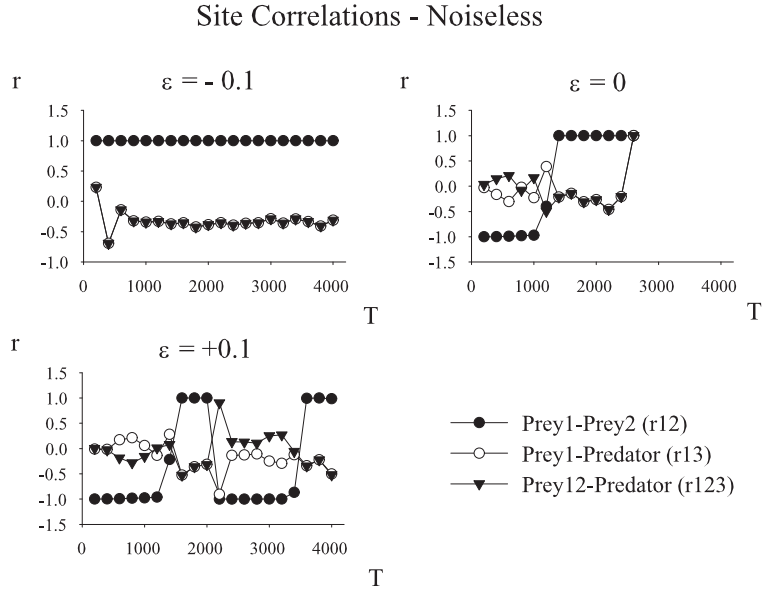}
\caption{Site correlation parameter r in noiseless dynamics as a
function of time for different values of the parameter $\epsilon:
- 0.1, 0.,+ 0.1$. Here $\eta = 0.2$. The values of the other
parameters are the same used for Fig.\ref{pat1}.}
  \label{cor1}
 \end{center}
\end{figure}
In the absence of noise and with constant interaction parameter
$\beta$ we obtain: (i) for $\epsilon < 0$ ($\beta < 1$) a
coexistence regime of the two preys characterized in the lattice
by a strong correlation between them with the predator lightly
anti-correlated with the two preys; (ii) for $\epsilon > 0$
($\beta > 1$) wide exclusion zones in the lattice (see
Fig.\ref{pat1}), characterized by a strong anti-correlation
between preys. In Fig.\ref{pat1} we report the spatial
configuration of the species after long time evolution. We chose
Gaussian initial distribution with mean value $ \bar{x}_o =
\bar{y}_o = \bar{z}_o = 0.25$ and variance $\sigma_o = 0.1$.
\begin{figure}[htbp]
 \begin{center}
  \includegraphics[height=5cm]{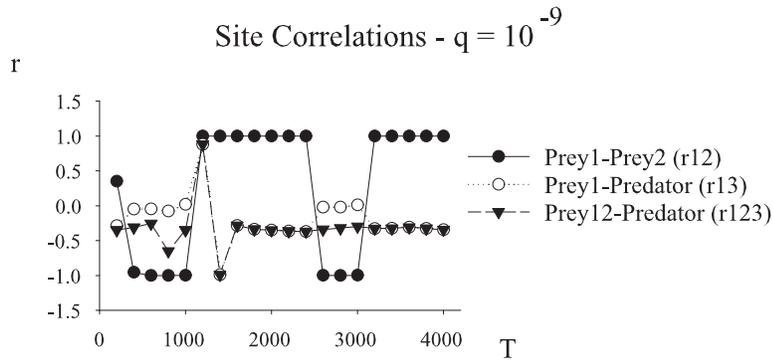}
\caption{Site correlation parameter as a function of time. Here
$\epsilon = -0.1, \eta = 0.2$ and the noise intensity is $q =
10^{-9}.$ The values of the other parameters are the same used for
Fig.\ref{pat1}. The initial spatial distribution is homogeneous
and equal for all species, i. e.
$x_{ij}^{init}=y_{ij}^{init}=z_{ij}^{init}= 0.25$ for all sites
($i,j$).}
  \label{cor2}
 \end{center}%
\end{figure}

By considering the periodic variation of the interaction parameter
$\beta(t)$, we obtain for $ \epsilon = 0 $, after a transient
anti-correlated behavior between preys, a coexistence regime with
strong correlation between preys that evolves towards an
homogeneous spatial distribution of all three species. For
$\epsilon > 0$ we find an oscillating behavior of the site
correlation coefficient from coexistence regime between preys,
corresponding to strong correlation, to an exclusion regime,
corresponding to strong anticorrelation. This last behavior is
prevalent. The oscillating frequency coincides with that of the
$\beta$-parameter. When $\epsilon < 0$, the two preys, after an
initial transient, remain strongly correlated for all the time, in
spite of the fact that the parameter $\beta(t)$ takes values
greater than $1$ during the periodical evolution. This situation
corresponds to a coexistence regime between preys. In
Fig.\ref{cor1} we report the behavior of the site correlation
parameter $r$ as a function of time for three values of the
parameter $\epsilon=-0.1, 0, 0.1$.

\begin{figure}[htbp]
 \begin{center}
  \includegraphics[height=4cm]{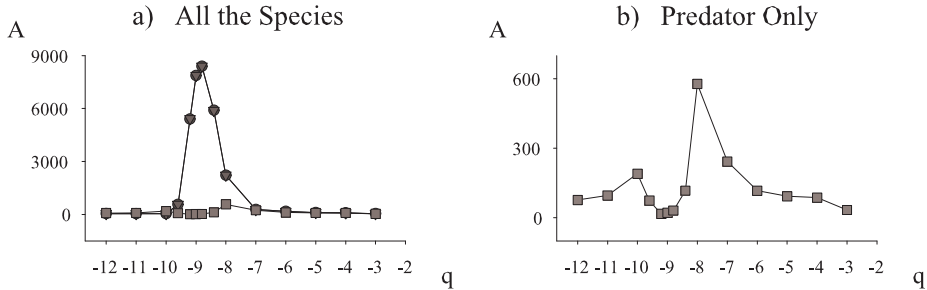}
\caption{Semi-Log plot of the mean area of the maximum patterns
for all species as a function of noise intensity, at iteration
step $1400$. Here circles and triangles are related to preys,
squares to predator. The values of the other parameters and the
homogeneous initial distribution are the same used in
Fig.\ref{cor2}.}
  \label{ar1}
 \end{center}%
\end{figure}

\subsection{Noise effects}

To analyze the effect of the noise we focus on the interesting
dynamical regime characterized, in absence of noise, by
coexistence between preys in all the period of $\beta$, i. e. with
$\epsilon<0$. The noise triggers the oscillating behavior of the
site correlation $r$ giving rise to periodical alternation of
coexistence and exclusion regime. Even a very small amount of
noise is able to destroy the coexistence regime periodically as we
can see from Fig.\ref{cor2}, where the correlation parameter $r$
as a function of time is reported.
\begin{figure}[htbp]
 \begin{center}
  \includegraphics[height=14cm]{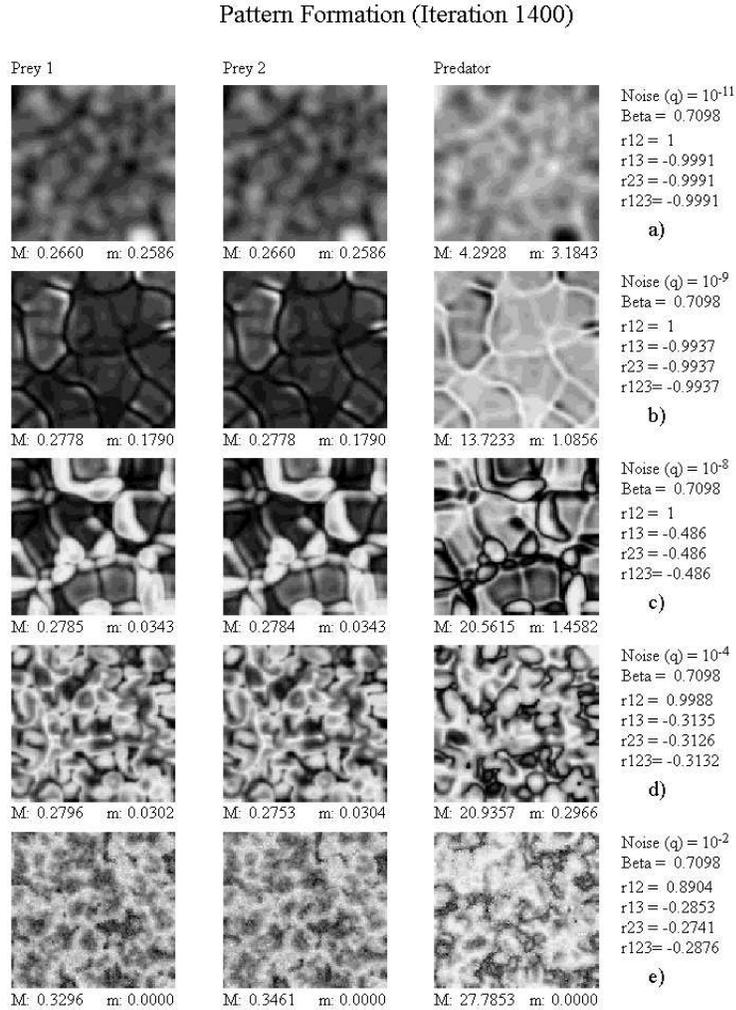}
\caption{Spatial Pattern formation for preys and predator, at time
iteration $1400$ and for the following values of the noise
intensity: $q = 10^{-11}, 10^{-9},10^{-8},10^{-4},10^{-2}$. The
values of the other parameters and the homogeneous initial
distribution are the same used in Fig.\ref{cor2}. The parameters
$r_{12}, r_{13},r_{23},r_{123}$ have the same meaning of
Fig.\ref{pat1}.}
  \label{pat2}
 \end{center}%
\end{figure}
Noise is also responsible for a nonmonotonic behavior of the area
of spatial patterns, which repeats periodically in time. In
Fig.\ref{ar1} we report a nonmonotonic behavior of the area of the
maximum pattern as a function of noise intensity. A maximum of the
area of maximum patterns is visible for the preys at $q = 10^{-9}$
and for the predator at $q = 10^{-8}$. The same behavior is
present in the following time steps within the first period of the
interaction parameter: 600, 800, 1200, 1400. But at time steps
$600, 800$ the preys are highly anticorrelated with site
correlation parameter $r_{12} = -1$, while at time steps
$1200,1400$ the preys are highly correlated with $r_{12} = 1$. The
formation of spatial patterns appear only when the preys are
highly correlated, while large patches with clusterization of
preys appear when they are anticorrelated. This means that the
coexistence regime between preys corresponds to the appearance of
spatial patterns, while the exclusion regime corresponds to
clusterization of preys. The noise-induced pattern formation
relative to the iteration 1400 is visible in Fig.\ref{pat2}, where
we report five patterns of the preys and the predator for the
following values of noise intensity: $q=10^{-11},
10^{-9},10^{-8},10^{-4},10^{-2}$. The initial spatial distribution
is homogeneous and equal for all species, that is
$x_{ij}^{init}=y_{ij}^{init}=z_{ij}^{init}= 0.25$ for all sites
($i,j$). A spatial structure emerges with increasing noise
intensity. At very low noise intensity, with respect to the value
of the diffusion coefficient $D$, the spatial distribution appears
almost homogeneous without any particular structure (see
Fig.\ref{pat2}a). Increasing the multiplicative noise intensity,
the symmetric dynamical evolution of the two preys in each site of
the lattice is destroyed, so oscillations in population density
produce an exclusion regime of one of two preys \cite{valenti}.
This time evolution scenario corresponds to the appearance of
spatial patterns due to different spatial density in each site.
This spatial pattern disappears for sufficiently large noise
intensity, producing a random spatial unhomogeneity (see
Fig.\ref{pat2}e). As a final investigation we analyze the behavior
of the area of the patterns as a function of time. We observe a
nonmonotonic behavior of the area of MPs as a function of time for
all values of the noise intensity investigated. Particularly for
noise intensity values greater than $q = 10^{-7}$ this
nonmonotonic behavior becomes periodical in time with the same
period of $\beta(t)$, as shown in Fig.\ref{graf} for $q =
10^{-4}$. We note that this nonmonotonic behavior doesn't mean
that a spatial pattern appears, like that of Fig.\ref{pat2}b, but
that a big clusterization of preys may occur, as shown in
Fig.\ref{pat3}. In this figure in fact we report the spatial
configuration obtained for a noise intensity value $q = 10^{-4}$,
which corresponds to the maximum of the behavior of the mean area
of MPs of preys as a function of time shown in Fig.\ref{graf}a. We
see that this maximum corresponds to large patches of preys in the
lattice investigated. The various quantities, such as pattern area
and correlation parameter, have been averaged over 200
realizations, obtaining the mean values shown in the figures
(3),(4) and (6). The effects induced by the interaction between
the species and the environment, modelled by the multiplicative
noise, can be summarized as: (i) to break the symmetry of the
coexistence regime between the preys, producing an alternation
with the exclusion regime; (ii) to trigger the oscillating
behavior of the site correlation coefficient; (iii) to produce a
nonmonotonic behavior of the pattern area as a function of the
noise intensity with an appearance of spatial patterns.

\begin{figure}[htbp]
 \begin{center}
  \includegraphics[height=4cm]{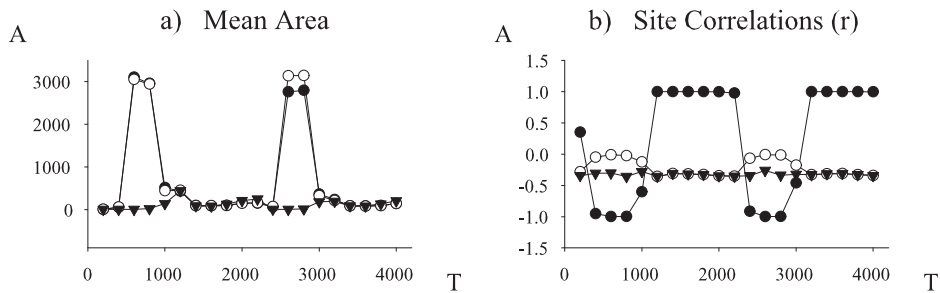}
\caption{Mean area of Maximum pattern of the three species and
relative sites correlations between preys and between preys and
predator as a function of time and $q = 10^{-4}$. (a): black and
white circles are related to preys, triangles to predator; (b)
site correlation $r_{12}$ (black circles), $r_{13}$ (white
circles), and $r_{123}$ (triangles). The values of the other
parameters and the homogeneous initial distribution are the same
used in Fig.\ref{cor2}.}
  \label{graf}
 \end{center}%
\end{figure}

\begin{figure}[htbp]
 \begin{center}
  \includegraphics[height=3cm]{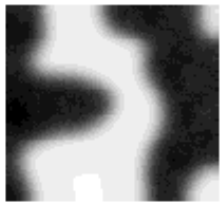}
\caption{Spatial configuration of one prey obtained for
$q=10^{-4}$, at iteration step $800$, corresponding to the maximum
value of the mean area for preys reported in Fig.\ref{graf}. The
values of the other parameters and the homogeneous initial
distribution are the same used in Fig.\ref{cor2}.}
  \label{pat3}
 \end{center}%
\end{figure}

\section{Conclusions}
The noise-induced pattern formation in a lattice of three
interacting species, described by Lotka-Volterra generalized
equations, has been investigated. We find nonmonotonic behavior of
the mean area of the maximum patterns as a function of noise
intensity. The same behavior we find for the area of the patterns
as a function of evolution time. The noise changes the dynamical
regime of the species, breaking the symmetry of the coexistence
regime. Besides the noise produces spatial patterns and temporal
oscillations of the site correlation parameter defined on the
lattice. We finally note that our simple model of an ecosystem of
interacting species could be useful to interpret the experimental
data of population dynamics strongly affected by the noise
\cite{Maz}.

\section{Acknowledgments}
 This work was
supported by \mbox{INTAS Grant 01-0450}, by INFM and MIUR.

\end{document}